\documentstyle[prb,aps]{revtex}
\begin{document}
\draft
\large
\title{Matching fields of a long superconducting film}
\author{Edson Sardella$^{\rm a}$, M.\ M.\ Doria$^{b}$, and P.\ R.\ S.\ Netto$^{\rm a}$}
\address{$^{\rm a}$Departamento de F\'{\i}sica, Universidade Estadual Paulista \\
Caixa Postal 473, 17033-360, Bauru-SP, Brazil}
\address{$^{\rm b}$Instituto de F\'{\i}sica, Universidade Federal do Rio de Janeiro\\
Caixa Postal 68528, 21945-970, Rio de Janeiro-RJ, Brazil}
\date{\today}
\maketitle
\begin{abstract}
We obtain  the vortex configurations, the matching fields and the magnetization of a superconducting film with a  finite cross section.
The applied magnetic field is normal to this cross section, and we
use London theory to calculate many of its properties, such as 
the local magnetic field, the free energy and the induction for the mixed state.
Thus previous similar theoretical works, done for an infinitely long superconducting film, are recovered here, in the special limit of a very long cross section.
\end{abstract}


\section{Introduction}\label{sec1}

The superconducting properties of an infinite type-II film have
been studied for more than three decades. As long as 1964, it was
first showed by Abrikosov \cite{abrikosov64} that, when an infinitely long film
is placed in an external magnetic field, parallel to its
surfaces,  vortices penetrate collectively into the sample in the
form of linear chains. 
He also calculated the lower critical field
$H_{c1}(b)$ as a function of the film thickness $b$ and
showed that for small $b$, it becomes increasingly more difficult for
a linear chain of vortices to penetrate the film.
After the work of Abrikosov, many
theoretical\cite{abrikosov64,shmidt,carter69,mawatari94,carneiro98} and
experimental
works\cite{brongersma93,brongersma94a,brongersma94b,guimpel87,hunnekes87}
have been carried out to study the mixed state of a type-II superconducting
film.
As the external magnetic field exceeds the value
of  $H_{c1}(b)$, the linear chains of vortices start to penetrate into
the sample, one at the time.
This reflects into the magnetization, which  exibits a succession of peaks related to the penetration of each linear chain.
The transition between two vortex states, differing by a linear chain,
occurs at a well defined critical magnetic field, known
as the {\it matching field}.

In this paper we determine the
matching fields and the magnetization of a film of finite cross
section, namely, with a finite width $a$ and also with a finite thickness $b$.
The applied field is always normal to this cross section, so that the third dimension is also taken very long here.
For $b$ very large we recover the results obtained by the previous authors in the
limit of an infinitely long film.
Interestingly there are two kinds of matching fields in this geometry, 
because  vortices penetrate individually here.
The first kind corresponds to 
the critical field for the penetration of an individual vortex. 
Thus there are many of this critical
field, each   delimiting the transition from a $N-$ to a $(N+1)-$ vortex
state. 
The second kind is  associated to a transition between different chain states, like in the case of an infinitely long film.
As vortices enter the finite film, they align themselves and form  a linear chain, parallel to  the longer side of the film. 
As the number of vortices increase inside the
sample, the linear chain  saturates and undergoes a transition,
splitting into two new chains. This process continues from two to three chains and so on.
The the critical fields associated to these transitions correspond to that required to enter a new linear chain in the case of the infinitely long film.

In summary we show here several new properties of the finite film, such as in 
the  formation of the equilibrium vortex state. 
The source of these new effects lays in the strong and very important
border effects that do not exist in case of the infinitely long film. 
For instance, a vortex state made of parallel chains is substantially
affected near the edges of the longer side of the film. 
In order to allow for any possible configuration,  we let the positions of the vortices entirely  free and minimize the Gibbs free energy with respect to them using a numerical procedure.
For this end, we use the Monte Carlo Simulated Annealing
minimization method. 
This method allows us to search for  the
global minimum of the Gibbs free energy (or at least very close to the global minimum) of the vortex system, which
corresponds to the most stable configuration.
We also solve here  for the matching 
fields, which  requires previous knowledge of the Gibbs free energy minimimum for a fixed number of vortices.
Notice that this program is much simpler for  an
infinite film, because there one can suppose some type of symmetry for 
the vortex state, and minimization can be reduced to a  small set of parameters, such as the lattice spacing and the number of chains\cite{carneiro98}.

Our analysis is restricted to the low magnetic
field regime and to the very strong type-II superconductors, where the
vortex-cores do not overlap and the distance over which the
magnetic field penetrates into the sample is much  larger than
the size of the cores. 
Within this regime, the mixed state
may be well described by the London theory accompanied by
some regularization procedure to cure some divergencies which
appear at short length scales. 
To the best knowledge of the present authors this study has not been done before.

\section{The Mixed State}\label{sec2}

The description of the mixed state of a type-II superconductor
involves two fundamental lengths. One is the distance over
which the magnetic field penetrates into the sample, the penetration
length $\lambda$. The other one is the coherence length, the
size of the vortex-core. Our treatment will be limited to the
case of a strong type-II superconductor where the Ginzburg-Landau
parameter $\kappa=\lambda/\xi\gg 1$. 
Vortices interact strongly since the distance between them can be much smaller than $\lambda$.
Their  repulsive interaction forbids overlapping, namely the onset
of vortices with multiple vorticity.
In this context, the
distribution of the local magnetic field may be well described by
the London equation. In rectangular coordenates $(x,y)$ the
London equation for the local magnetic field ${\bf h}=h{\bf z}$
is given by

\begin{equation}
-\lambda^2\nabla^2h+h=\Phi_0\,\sum_i\,\delta ({\bf r}-{\bf r}_i)\;,
\label{le}
\end{equation}
where $\Phi_0$ is the quantum flux and ${\bf r}_i=(x_i,y_i)$ is
the position of the $i$-th vortex. We have assumed that the
vortices are straight lines. Therefore, a 3D problem is
reduced to a 2D one.

The geometry of the problem is ilustrated in Fig. \ref{fig1}. Let
$a$ and $b$ be the sides of the superconducting film. We will
solve this equation subject to the following boundary conditions

\begin{eqnarray}
& & h(\pm a/2,y)=h(x,0)\;=\;h(x,b)\;=H\;, \nonumber \\
& & \left ( \frac{\partial h}{\partial x}\right )_{y=0,b}=\left (
\frac{\partial h}{\partial y}\right )_{y=\pm a/2}\;=\;0\;,
\label{bch}
\end{eqnarray}
where $H$ is the magnitude of the external magnetic field at the interface
vacuum-superconductor and is pointing along the $z$-direction. The second
condition imposes that the
supercurrents are confined into the sample.

Other authors have used the image method to solve the London
equation (See for instance Ref. \onlinecite{carter69} 
and \onlinecite{mawatari94}). We prefer to solve this equation 
directly, by using the Green's function method. The equation for 
this function associated with the London equation is given by

\begin{equation}
-\lambda^2\nabla^2G+G=\delta(x-x^{\prime})\delta(y-y^{\prime})\;,
\label{ge}
\end{equation}
where $G$ must satisfy the following boundary conditions on the
film borders

\begin{equation}
G(\pm a/2,y)=G(x,0)=G(x,b)=0\;.\label{bcg}
\end{equation}

We can find an expression for the local magnetic field in terms
of the Green's function by multipling (\ref{le})
by $G(x,y,x^{\prime},y^{\prime})$ and (\ref{ge}) by $h(x,y)$,
subtract the results and then integrate what is left. One obtains

\begin{eqnarray}
h(x^{\prime},y^{\prime}) & = & \phi_0\,\sum_i\,G(x_i,y_i,x^{\prime},y^{\prime})
\nonumber \\
& & -H\left [ 1-\int_{-a/2}^{a/2}\,dx\,\int_{0}^{b}\,dy\,
G(x,y,x^{\prime},y^{\prime})\right ]\;,\label{hge}
\end{eqnarray}
where we have made use of the boundary conditions
(\ref{bch}) and (\ref{bcg}).
 
Then, the solution for the local magnetic field is transfered to the
determination of the Green's function. The approach used to
find the solution for this function is sketched in full
details in Ref. \onlinecite{jackson}, except by the fact
that there the boundary conditions are taken at infinite. Here
we only show the main steps of how to calculate this
function. First of all we expand $G$ in a Fourier series

\begin{equation}
G(x,y,x^{\prime},y^{\prime})=\frac{2}{b}\,\sum_{m=1}^{\infty}\,
\sin\left ( \frac{m\pi y^{\prime}}{b}\right )
\sin\left ( \frac{m\pi y}{b}\right )g_m(x,x^{\prime})\;,\label{gf}
\end{equation}
which satisfies the boundary conditions of (\ref{bcg}) at $y=0,b$.

Inserting this equation into (\ref{ge}) we find for $g_m$

\begin{equation}
-\lambda^2 \frac{\partial^2 g_m}{\partial x^2}+\alpha_m^2g_m=
\delta(x-x^{\prime})\;,\label{gme}
\end{equation}
where we have used the fact that the sequence
$\left \{ \sqrt{\frac{2}{b}}\,\sin\left ( \frac{m\pi y}{b}
\right )\;m=1,2,3\ldots \right \}$ is a complete set of
orthonormal functions, that is,

\begin{equation}
\frac{2}{b}\,\sum_{m=1}^{\infty}\,\sin\left ( \frac{m\pi
y^{\prime}}{b}\right )
\sin\left ( \frac{m\pi y}{b}\right )=\delta(y-y^{\prime})\;.
\end{equation}

Here

\begin{equation}
\alpha_m=\left [ 1+\lambda^2\left ( \frac{m\pi}{b}\right )^2
\right ]^{1/2}\;.
\end{equation}

The function $g_m(x,x^{\prime})$ must satisfy the same boundary
conditions as $G(x,y,x^{\prime},y^{\prime})$, that is,
$g_m(\pm a/2,x^{\prime})=0$. In addition, the derivative of
$g_m(x,x^{\prime})$ is discontinous at $x=x^{\prime}$ which
results from the delta function on the right-hand side of (\ref{gme}).

Under these conditions, the solution for $g_m(x,x^{\prime})$ is given by

\begin{equation}
g_m(x,x^{\prime})=\frac{1}{2\lambda\alpha_m\sinh (\alpha_m a/\lambda)}
\left \{ \cosh [ \alpha_m ( |x-x^{\prime}|-a)/\lambda ]-\cosh [
\alpha_m ( x+x^{\prime})/\lambda ]\right \}\;.\label{spgm}
\end{equation}

This completes the solution for the Green's function. To proceed we still
have to find the integral of the Green's function which appears in
(\ref{hge}). This involves a tedious, but straightforward algebra. One has

\begin{eqnarray}
\int_{-a/2}^{a/2}\,dx\,\int_{0}^{b}\,dy\,
G(x,y,x^{\prime},y^{\prime}) & = & \frac{4}{b}\,\sum_{m=1}^{\infty}\,
\frac{(-1)^m}{\alpha_{2m+1}^2}\frac{b}{(2m+1)\pi}
\sin\left ( \frac{m\pi y^{\prime}}{b}\right )\nonumber \\ 
& & \times\left [
1-\frac{\cosh (\alpha_{2m+1} x^{\prime}/\lambda)}{\cosh (\alpha_{2m+1}
a/2\lambda)}
\right ]\;.\label{ifg}
\end{eqnarray}

Notice that this vanishes at $x^{\prime}=\pm a/2$ or 
$y^{\prime}=0,b$, so that the second term of (\ref{hge}) 
is just $H$ as required by the boundary conditions of (\ref{bch}).

Next we can simplify (\ref{ifg}) by noticing that

\begin{eqnarray}
\frac{4}{b}\,\sum_{m=1}^{\infty}\,
\frac{(-1)^m}{\alpha_{2m+1}^2}\frac{b}{(2m+1)\pi}
\sin\left ( \frac{m\pi y^{\prime}}{b}\right ) & = & 
\int_{0}^{b}\,dy\,g(y,y^{\prime})\nonumber \\
& = & 1-\frac{\cosh [(y^{\prime}-b/2)/\lambda]}{\cosh (b/2\lambda)}\;,
\label{ifgu}
\end{eqnarray}
where

\begin{eqnarray}
g(y,y^{\prime}) & = & \frac{2}{b}\,\sum_{m=1}^{\infty}\,\frac{1}{\alpha_m^2}
\sin\left ( \frac{m\pi y^{\prime}}{b}\right )
\sin\left ( \frac{m\pi y}{b}\right )\nonumber \\
& = & \frac{1}{2\lambda\sinh (b/\lambda)}
\left \{ \cosh [ (|y-y^{\prime}|-b)/\lambda ]-\cosh [
(y+y^{\prime})/\lambda ]\right \}\;.
\end{eqnarray}

By combining (\ref{hge}), (\ref{ifg}) and (\ref{ifgu}) we finally obtain for
the local magnetic field

\begin{eqnarray}
h(x,y) & = & \Phi_0\,\sum_i\,G(x,y,x_i,y_i)+H\left \{ 
\frac{\cosh [(y-b/2)/\lambda]}{\cosh(b/2\lambda)}\right . \nonumber \\
& & +\left . \frac{4}{b}\,\sum_{m=0}^{\infty}\,
\frac{(-1)^m}{\alpha_{2m+1}^2}\frac{b}{(2m+1)\pi}
\sin\left [ \frac{(2m+1)\pi y}{b}\right ]\frac{\cosh (\alpha_{2m+1}
x/\lambda)}{\cosh (\alpha_{2m+1} a/2\lambda)}\right \} \;,\label{sph}
\end{eqnarray}
where we have made use of the following symmetry property of the Green's 
function, $G(x,y,x^{\prime},y^{\prime})=G(x^{\prime},y^{\prime},x,y)$. We 
can see that the local magnetic field is composed essentially by three 
contributions. The last two terms, proportional to $H$ in (\ref{sph}),
represent the penetration of the external magnetic field near 
the surface when the vortices are still absent. Once the vortices 
start penetrating the sample they will be trapped by the shielding 
currents associated with the last two terms of (\ref{sph}). The Green's
function in (\ref{sph}) contains two terms (see (\ref{gf}) and
(\ref{spgm})). The first one, which depends of $|x-x_i|$, represents the
local magnetic field of the vortices located inside the sample, and the
second one, which depends on $x+x_i$ is the local magnetic 
field of the image vortices located outside the sample.

The free energy per unit volume of the superconducting film is given by

\begin{eqnarray}
{\cal F} & = & \frac{1}{8\pi A}\,\int\,d^2r\,\left \{ h^2+
\lambda^2\left [ \left ( \frac{\partial h}{\partial x}\right )^2
+\left ( \frac{\partial h}{\partial y}\right )^2\right ]\right \}
\nonumber \\
& = & \frac{\Phi_0}{8\pi A}\,\sum_i\,h(x_i,y_i)+\frac{H\lambda^2}{8\pi
A}\left \{ \int_{0}^{b}\,dy\,\left [ 
\left ( \frac{\partial h}{\partial x}\right )_{x=a/2}-
\left ( \frac{\partial h}{\partial x}\right )_{x=-a/2}\right ] 
\right . \nonumber \\
& & \left . +\int_{-a/2}^{a/2}\,dx\,\left [ 
\left ( \frac{\partial h}{\partial y}\right )_{y=b}-
\left ( \frac{\partial h}{\partial y}\right )_{y=0}\right ]
\right \}\;,\label{len}
\end{eqnarray}
where $A=ab$.

Now notice that by integrating the kinetic term of the London equation 
of (\ref{le}) we obtain precisely the second term of (\ref{len}). Then, 
this equation can be simplified to

\begin{equation}
{\cal F}=\frac{\Phi_0}{8\pi A}\,\sum_i\,h(x_i,y_i)+\frac{HB}{8\pi}
-N\frac{\Phi_0H}{8\pi A}\;,\label{lenb}
\end{equation}
where $N$ is the number of vortices, $B$ is the spatial average of the 
local magnetic field. The induction is obtained by integrating (\ref{sph})
over the cross section of the superconducting film. One has

\begin{eqnarray}
AB & = & N\Phi_0-\Phi_0\,\sum_i\,\frac{\cosh (y_i-b/2)/\lambda)}{\cosh
(b/2\lambda)} 
\nonumber \\
& & -\frac{4\Phi_0}{\pi}\,\sum_i\,\sum_{m=0}^{\infty}\,\frac{(-1)^m}
{\alpha_{2m+1}^2}\frac{1}{(2m+1)}
\sin\left [ \frac{(2m+1)\pi y_i}{b}\right ]\frac{\cosh (\alpha_{2m+1}
x_i/\lambda)}{\cosh (\alpha_{2m+1} a/2\lambda)} \nonumber \\
& & +HA\left \{ \frac{\tanh (b/2\lambda)}{(b/2\lambda)}-\frac{8}{\pi^2}\,
\sum_{m=0}^{\infty}\,\left [ \frac{1}{(2m+1)\alpha_{2m+1}}\right ]^2
\frac{\tanh (\alpha_{2m+1} a/2\lambda)}{(\alpha_{2m+1} a/2\lambda)}
\right \}\;.\label{ind}
\end{eqnarray}

Upon substituting (\ref{ind}) into (\ref{lenb}) we are left with

\begin{eqnarray}
{\cal F} & = & \frac{\Phi_0^2}{8\pi
A}\,\sum_{i,j}\,G(x_i,y_i,x_j,y_j)\nonumber \\
& & +\frac{H^2}{8\pi}\left \{ \frac{\tanh
(b/2\lambda)}{(b/2\lambda)}-\frac{8}{\pi^ 2}\,
\sum_{m=0}^{\infty}\,\left [ \frac{1}{(2m+1)\alpha_{2m+1}}\right ]^2
\frac{\tanh (\alpha_{2m+1} a/2\lambda)}{(\alpha_{2m+1} a/2\lambda)}
\right \}\;.\label{lenf}
\end{eqnarray}

In order to study the most stable configuration of the vortex 
lattice, it is more convenient to take the Gibbs free energy. 
In units of volume, its expression is given by ${\cal G}={\cal F}-BH/4\pi$. 
From (\ref{ind}) and (\ref{lenf}), it follows that

\begin{eqnarray}
{\cal G} & = & \frac{\Phi_0^2}{8\pi A}\,\sum_{i,j}\,G(x_i,y_i,x_j,y_j)
+\frac{\Phi_0H}{4\pi A}\,\sum_i\,\frac{\cosh [ (y_i-b/2)/\lambda ]}
{\cosh (b/2\lambda)}\nonumber \\
& & +\frac{\Phi_0H}{\pi A}\,\sum_i\,\sum_{m=0}^{\infty}\,\frac{(-1)^m}
{\alpha_{2m+1}^2}\frac{1}{(2m+1)}
\sin\left [ \frac{(2m+1)\pi y_i}{b}\right ]\frac{\cosh (\alpha_{2m+1}
x_i/\lambda)}{\cosh (\alpha_{2m+1} a/2\lambda)}\nonumber \\
& & -\frac{H^2}{8\pi}\left \{ \frac{\tanh
(b/2\lambda)}{(b/2\lambda)}-\frac{8}{\pi^2}\,
\sum_{m=0}^{\infty}\,\left [ \frac{1}{(2m+1)\alpha_{2m+1}}\right ]^2
\frac{\tanh (\alpha_{2m+1} a/2\lambda)}{(\alpha_{2m+1} a/2\lambda)}
\right \} \nonumber \\
& & -N\frac{\Phi_0H}{4\pi A}\;.\label{gfen}
\end{eqnarray}

The last two terms are the energy of the Meissner state as if 
no vortices were present. The second and third contributions are 
the potential barrier which pin the vortices inside the sample. 
The Green's function in the first term gives rise to two contributions. 
One is the repulsive interaction between the vortices and the other 
one is the atractive interaction between the vortices and 
the images which are virtually placed outside the sample.

We shall work in the limit of thin film in which $(\pi\lambda/b)^2\gg 1$.
Within this limit, the sum in $m$ can be evaluated exactly. We proceed as in
Ref. \onlinecite{mawatari94}. One has,

\begin{eqnarray}
G(x_i,y_i,x_j,y_j) & = & \frac{1}{4\pi\lambda^2}\left \{ \ln
\left [ 
\frac{\cosh [\pi |x_i-x_j|/b]-\cos [\pi (y_i+y_j)/b]}
{\cosh [\pi |x_i-x_j|/b]-\cos [\pi (y_i-y_j)/b]} 
\right ] \right . \nonumber \\
& & \left . -\ln
\left [
\frac{\cosh [\pi (a-x_i-x_j)/b]-\cos[\pi (y_i+y_j)/b]}
{\cosh [\pi (a-x_i-x_j)/b]-\cos[\pi (y_i-y_j)/b]}
\right ] \right . \nonumber \\
& & \left . -\ln
\left [
\frac{\cosh [\pi (a+x_i+x_j)/b]-\cos[\pi (y_i+y_j)/b]}
{\cosh [\pi (a+x_i+x_j)/b]-\cos[\pi (y_i-y_j)/b]}
\right ]
\right \}\;.\label{gapprox}
\end{eqnarray}

The first term in (\ref{gapprox}) is the same as the one found in Ref.
\onlinecite{mawatari94}. The other two terms are due to the presence of the
lateral border of the film. London theory 
is not regular for vortex self-interaction. In fact, one 
can notice that the contribution $i=j$ to the Gibbs free energy (\ref{gfen})
gives rise to a logarithmic divergence. 
We remediate this divergence by using a sharp cutoff in which $|x_i-x_j|$ is
replaced by $\xi$ for $i=j$. In the limit $(b/\pi\xi)\gg 1$ we obtain

\begin{eqnarray}
G(x_i,y_i) & = & \frac{1}{4\pi\lambda^2}\left \{ \ln
\left [ 
\frac{(\pi\xi/b)^2+4\sin^2(\pi y_i/b)}
{(\pi\xi/b)^2} 
\right ] \right . \nonumber \\
& & \left . -\ln
\left [
\frac{\cosh [\pi (a-2x_i)/b]-\cos(2\pi y_i/b)}
{\cosh [\pi (a-2x_i)/b]-\cos(\pi\xi/b)}
\right ] \right . \nonumber \\
& & \left . -\ln
\left [
\frac{\cosh [\pi (a+2x_i)/b]-\cos(2\pi y_i/b)}
{\cosh [\pi (a+2x_i)/b]-\cos(\pi\xi/b)}
\right ]
\right \}\;.\label{gapprox_self}
\end{eqnarray}

The sum over $m$ in the third term on the right-hand side of (\ref{lenf})
cannot be evaluated exatly in the limit $(\pi\lambda/b)^2\gg 1$. So it will
be kept as it stands. 

\section{Results and Discussion}\label{sec3}

The lower critical field, the magnitude of the external field sufficient to 
create a vortex inside the sample, may be obtained by neglecting the 
quadratic term in (\ref{gfen}) and equating the remaining terms 
to zero. We find

\begin{equation}
H_{c1}(a,b)=\frac{\Phi_0}{2\lambda^2}\left \{ \frac{\frac{1}{2\pi}\ln\left (
\frac{2b}{\pi\xi}
\right )
-
\ln\left [
\frac{\cosh (\pi a/b)+1}{\cosh (\pi a/b)-1}
\right ]
}{1-\frac{1}{\cosh (b/2\lambda)}-4\,\sum_{m=0}^{\infty}\,
\frac{(-1)^m}{(2m+1)\pi\alpha_{2m+1}^2}\frac{1}{\cosh (\alpha_{2m+1}
a/2\lambda)}
}\right \}\;,\label{hc1}
\end{equation}
where we have assumed that a single vortex is at the center of the film. It
can be easily seen that for $a\rightarrow\infty$ we obtain the same result
as in Ref. \onlinecite{carneiro98}. A quick inspection of (\ref{hc1}) shows
us that as $a$ and/or $b$ decreases, $H_{c1}(a,b)$ increases. In other words,
size effects provoke a delay in the first penetration of flux, assuming the external field increasing with time.
These 
finite size effects have been observed in supercondutors of different
geometry \cite{venegas98}. The determination of the other critical fields
will be discussed below. 

Now we turn our discussion to the minimization of the Gibbs free energy 
with respect to the vortex positions. Our investigation will be guided 
by the Monte Carlo Simulated Annealing minimization method. We have 
done this from $N=1$ up to $N=30$. We start from an arbitrary 
initial configuration of the vortex positions at a certain ``temperature''.
By adiabatically 
lowering this ficticious temperature, Simulated Annealing will 
concentrate on the region where it is most likely to succeed.

The parameters we use here are $\kappa=100$, $a=7\lambda$, and
$b=0.75\lambda$. We find that, as the penetration of flux is initiated, the
vortices 
will be symmetrically located with respect to the center of the film 
in form of a linear chain aligned parallel to the $x$-axis (the longer side
of the film). As $H$ is increased, at a certain critical value of the
external magnetic field, this linear chain breaks into a double chain. At
the interior of the film, the vortices arrange in form of a ``zig-zag''.
However, at the edges of the film both linear chains are distorted and
joined together (see Fig. \ref{fig2}). For the parameters quoted above, we
have found that the change in the symmetry of the lattice occurs at $N=24$.
This critical value of $N$ is expected to grow in case the value of 
the width $a$ of the film becomes larger.

In Fig. \ref{fig2}, we also have shown some other configurations for higher
values of $N$. Notice that the vortex state develops some unusual 
patterns in which the linear chains entangle in the interior of the film.
On each 
step of our numerical calculation, to search the minimum $G$ we lowered the
temperature as slowly as possible. However, we cannot assure that the
minimum $G$ found corresponds to the global optimum. So, those entangled
chain configurations may be metastable solutions.

The penetration of each additional vortex corresponds to a well defined 
value of external magnetic field which we denote  by $H_{sN}$; 
$H_{s1}\equiv H_{c1}(a,b)$. We find these matching fields by assuming 
that at the transition from $N$ to $N+1$ vortices, the Gibbs free energy 
is continuous, that is, ${\cal G}_N={\cal G}_{N+1}$. Because 
the vortex positions depend of the external magnetic field, this is 
a trancendental equation. Full details 
of how to solve this transcendental equation can be found in Ref. 
\onlinecite{venegas98}. We have determined the sequence of matching fields
for the parameters quoted above. We then used these values to 
calculate the  magnetization, which is 
defined by $M=(B-H)/4\pi$. 

The magnetization as a function of $H$ 
is depicted in Fig. \ref{fig3}. 
At the transitions, the creation of another 
vortex will force a rearrangement in the chain. This in turn will 
provoke a discontinuity of the magnetization at $H=H_{sN}$. 
We can 
see that the magnetization changes in small steps. So, the magnetization can
be used to measure the number of individual vortex penetration in the
sample. Each peak indicates that a new vortex entered the film.

Notice that $-4\pi M$ has  a maximum for $H\geq H_{c1}(a,b)$, signaling a
transition from a single to a double chain state,
in constrast to  superconducting systems with no edges, that just show
a  monotonic behavior. 
In each phase between  $H_{sN}$ and $H_{s,N+1}$, for each value of $H$ we should calculate 
the corresponding equilibrium configuration of the vortices and use it to 
evaluate the induction $B$. For the parameters used in Fig. \ref{fig3}, 
the differences $\Delta H=H_{s,N+1}-H_{sN}$ are not to large. Then, the vortex 
positons should not vary significantly with $H$ within each 
phase. So, to calculate $B$, we held the vortex positions fixed at 
$H=H_{sN}$. As a consequence, the magnetization varies linearly 
with $H$ within each phase.

We can also observe that the intensities of the peak of the magnetization at
the transition from a single to a double chain lattice, and at $H=H_{c1}(a,b)$
have approximately the same intensity. In Fig. \ref{fig4} we repeated the
same calculation with $a=7\lambda$ and $b=0.5\lambda$. For $N$ up to 30
vortices we did not find any break in the linear chain (see Fig.
\ref{fig4}). However,  the magnetization curve shows a tendency to reach the
second peak of slower magnitude than at $H=H_{c1}$. On the other hand, for
an infinite film, as shown in Ref. \onlinecite{carneiro98}, the intensity of
the magnetization at the phase transitions tends to grow as $H$ increases.
We then suspect that this behavior of the magnetization may be a result of
size effects.

Finally, we would like to point out that the successive discontinuities
of the magnetization discussed here  should be experimentally
observed. 
To see this, we note that the order of magnitude of $\Delta
H=H_{s,N+1}-H_{sN}$ in Fig. \ref{fig3} can be as high as
$\Phi_0/2\lambda^2$. For zero temperature, $\lambda$ is typically of order
$10^5$ \AA which gives $\Delta H\sim 10^{-5}$ T. On the other hand, the jump
in the magnetization can achieve the magnitude of order $10^{-7}$ T. This is
well inside the resolution of a magnetometer.

\acknowledgements
ES thanks the Brazilinan Agencies FAPESP and CNPq for financial suport. PRSN
thanks CNPq for financial suport.

\begin{figure}
\caption{Cross section of the superconducting film. The external 
magnetic field is along the $z$-axis and its value is $H$ for $|x|>a/2$, and
$y<0$ and $y>b$.}\label{fig1}
\end{figure}

\begin{figure}
\caption{
The double chains for $N=24,25,26,27$, in the clockwise sense. The
parameters used are quoted in the text. Notice that the inner vortices make
a ``zig-zag'' as in a triangular lattice. However, at the edges of the
film, both chains end up at the same vortex.
}\label{fig2}
\end{figure}

\begin{figure}
\caption{The magnetization as a function of the external magnetic field $H$.
The parameters used are quoted in the text.}\label{fig3}
\end{figure}

\begin{figure}
\caption{The magnetization as a function of the external magnetic field $H$.
The parameters used are quoted in the text.}\label{fig4}
\end{figure}

\end{document}